\documentclass[12pt,twocolumn]{article}
\usepackage{tabularx}
\usepackage[english]{babel}
\usepackage[utf8x]{inputenc}
\usepackage[T1]{fontenc}

\usepackage{longtable}
\usepackage{setspace} 
\usepackage{multirow}

\usepackage{flushend}

\usepackage{enumitem}

\usepackage[twocolumn,top=1cm,bottom=1cm,left=1cm,right=1cm,marginparwidth=1cm]{geometry}
\usepackage{geometry}

\usepackage{amsmath}
\usepackage{graphicx}
\usepackage[colorinlistoftodos]{todonotes}
\usepackage[colorlinks=true, allcolors=blue]{hyperref}
\usepackage{authblk}

\providecommand{\keywords}[1]
{
  \immediate	
  \textbf{\textit{Keywords-}} #1
}

\title{Advances In Malware Detection-An Overview}

\date{}

\makeatletter
\newcommand\level[1]{%
  \ifcase#1\relax\expandafter\chapter\or
    \expandafter\section\or
    \expandafter\subsection\or
    \expandafter\subsubsection\else
    \def\next{\@level{#1}}\expandafter\next
  \fi}
\newcommand{\@level}[1]{%
  \@startsection{level#1}
    {#1}
    {\z@}%
    {-3.25ex\@plus -1ex \@minus -.2ex}%
    {1.5ex \@plus .2ex}%
    {\normalfont\normalsize\bfseries}}

\newcounter{level4}[subsubsection]
\@namedef{thelevel4}{\thesubsubsection.\arabic{level4}}
\@namedef{level4mark}#1{}
\count@=4
\loop\ifnum\count@<100
  \begingroup\edef\x{\endgroup
    \noexpand\newcounter{level\number\numexpr\count@+1\relax}[level\number\count@]
    \noexpand\@namedef{thelevel\number\numexpr\count@+1\relax}{%
      \noexpand\@nameuse{thelevel\number\count@}.\noexpand\arabic{level\number\numexpr\count@+1\relax}}
    \noexpand\@namedef{level\number\numexpr\count@+1\relax mark}####1{}}
  \x
  \advance\count@\@ne
\repeat
\makeatother
\author
{
Heena$^{a,b,c}$ , B. M. Mehtre$^{a,d}$  \\

\small $^{a}$Center of excellence in cyber security, Institute for Development and Research in Banking Technology (IDRBT), Hyderabad, India \\

\small $^{b}$School of Computer Science and Information Sciences
(SCIS), University of Hyderabad, Hyderabad, India\\ 

\small$^{c}$ heenarao014@gmail.com \\
\small$^{d}$ bmmehtre@idrbt.ac.in \\
        
} 

\begin{document}

\maketitle

\begin{abstract}
Malware has become a widely used means in cyber attacks in recent decades because of various new obfuscation techniques used by malwares. In order to protect the systems, data and information, detection of malware is needed as early as possible. There are various studies on malware detection techniques that have been done but there is no method which can detect the malware completely and make malware detection problematic. Static Malware analysis is very effective for known malwares but it does not work for zero day malware which leads to the need of dynamic malware detection and the behaviour based malware detection is comparatively good among all detection techniques like signature based, deep learning based, mobile/IOT and cloud based detection but still it is not able to detect all zero day malware which shows the malware detection is very challenging task and need more techniques for malware detection. This paper describes a literature review of various methods of malware detection. A short description of each method is provided and discusses various studies already done in the advanced malware detection field and their comparison based on the detection method used, accuracy and other parameters. Apart from this we will discuss various malware detection tools, dataset and their sources which can be used in further study. This paper gives you the detailed knowledge of advanced malwares, its detection methods, how you can protect your devices and data from malware attacks and it gives the comparison of different studies on malware detection. 
\end{abstract}

\keywords{Cyber Security, Malware Detection approaches, Malware Classification, Malware features.}

\section{Introduction}
Technology nowadays has become so advanced, everything is adapting the digital over the manual way of working. Technology has its pros and cons, if it makes life easier then at the same time it invites the cyber attacks, loss of data, giving access to your personal life to someone who can misuse it. So the security of our devices is very important in today’s cyber world. The internet usage is increasing day by day. One drawback of the widespread use of the internet is that many computer systems are vulnerable to attacks and get infected with malwares. There are different names for malware for example malicious code, malicious program or malicious executable. Malware is malicious software which is used with the intention of breaching a computer system’s security policy with respect to confidentiality, integrity and availability of data[33]. It can change and remove your system, data without your knowledge to harm the system. According to multiple studies the complete 100\% malware detection problem is NP complete problem [39][42], but in every study the researchers always try to get maximum accuracy by using different methods.

\subsection{Types of Malware}
\begin{figure}[htpb!] % Defines figure environment
    \centering % Centers your figure
\includegraphics[scale=0.5]{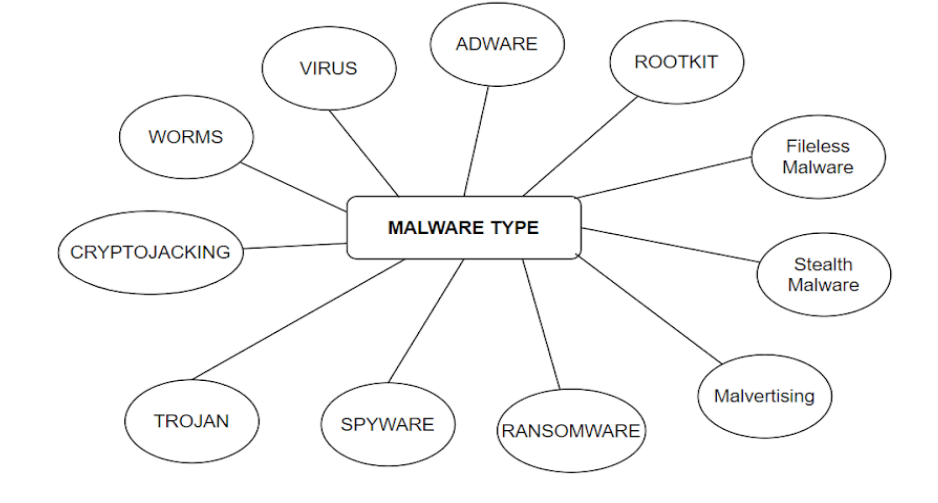} % Includes your figure and defines the size
    \caption{Types Of Malware} % For your caption
    \label{fig:label1} % If you want to label your figure for in-text references
\end{figure}

Figure~\ref{fig:label1} shows the different types of malware .
\subsubsection{ Virus}
Viruses attach their malicious code to clean code and wait for an unsuspecting user or an  automated process to execute them. Like a biological virus, they can spread quickly and widely, causing damage to the core functionality of systems, corrupting files and locking  users out of their computers. They usually hide within an executable file.[34].
\subsubsection{Worm}
Worms get their name from the way they infect systems. Starting from one infected  machine, they weave their way through the network, connecting to consecutive machines  in order to continue the spread of infection. This type of malware can infect entire networks of devices very quickly[34].
\subsubsection{Trojans}
This type of malware hides within or disguises itself as legitimate software. Acting  discreetly, it will breach security by creating backdoors that give other malware variants  easy access[34].
\subsubsection{Spyware}
It Hides in the background on a computer and this type of malware will collect information without the user knowing, such as credit card details, passwords and other sensitive information.\newline Spyware is software that is installed on your computer either directly or inadvertently. A Trojan horse program is similar to spyware except that it is packaged as another program[34].
\subsubsection{Ransomware}
Also known as scareware, ransomware comes with a heavy price. They are able to lockdown networks and lock out users until a ransom is paid, ransomware has targeted some of the biggest organizations in the world today with expensive results[43].
\subsubsection{Adware}
Adware is malware that forces your browser to redirect to web advertisements, which often  themselves seek to download further, even more malicious software[34].
\subsubsection{Rootkit}
Rootkit is, a program or, more often, a collection of software tools that gives a threat actor remote access to and control over a computer or other system. It gets its name because it's a kit of tools that (generally illicitly) gain root access (administrator-level control, in Unix terms) over the target  system, and use that power to hide their presence[43].
\subsubsection{Cryptojacking}
Cryptojacking is the unauthorized use of someone else's computer to mine cryptocurrency.\newline
Cryptojacking is another way attackers can force you to supply them with Bitcoin, only it works  without you necessarily knowing. The crypto mining malware infects your computer and uses your  CPU cycles to mine Bitcoin for your attacker's profit. The mining software may run in the  background on your operating system or even as JavaScript in a browser window[43].
\subsubsection{Malvertising}
Malvertising is the use of legitimate ads or ad networks to covertly deliver malware to  unsuspecting user’s computers. For example, a cybercriminal might pay to place an ad on a  legitimate website. When a user clicks on the ad, code in the ad either redirects them to a malicious website or installs malware on their computer. In some cases, the malware embedded in an ad might execute automatically without any action from the user, a technique referred to as a “drive by download”[43].
\subsubsection{Fileless}
Fileless malware attacks do not download malicious files or write any content to the disk in order to compromise the systems. The attacker exploits merely the vulnerable application to inject malicious code directly into the main memory. The attacker can also leverage the trusted and widely used applications, i.e., Microsoft office or administration tools native to Windows OS like PowerShell and WMI to run scripts and load malicious code directly into volatile memory[11].
\subsubsection{Stealth Malware}
A stealth virus is a hidden computer virus that attacks operating system processes and averts typical anti-virus or anti-malware scans. Stealth viruses hide in files, partitions and boot sectors and are adept at deliberately avoiding detection. Stealth malware are of different types based on what they are hiding. Stealth malware uses the hooking technique to divert the original system call to malware. There are mainly 4 types of stealth malware rootkits, code mutation, anti emulation, targeting mechanism[30]. Rootkit can use user mode hooking, kernel mode hooking or hybrid by combining both to inject malicious code[41]. Code mutation is malware changing its code to hide from antivirus using mutation engines but it can be detected via emulation. The other type of stealth malwares anti-emulation behaves differently while running in an emulated environment. They sense the environment and change the behavior according to the environment. The targeted mechanism of stealth malware runs and spreads only on the chosen systems. There are different countermeasures for component based and pattern based stealth malware. To detect component based stealth malware we can use the technique of detecting hooks using signature and heuristic methods which result in high false positive rate or Cross-View Detection and Specification Based Methods in which the output of API calls are compared with low-level calls that are designed to do the same thing.The other countermeasure is using hardware solution where  a clean machine can be use to monitor another machine for the presence of rootkits/stealth malware. Virtualization techniques are also able to detect stealth malware but these are also vulnerable to anti-mutation malware. For detecting pattern based stealth malware signature based, behaviour, heuristic and model based techniques can be used. There are multiple studies on stealth malware detection but no one gives the good result and need to find more good methods and studies[30][41][44][45].
\subsection{Malware Spreading Techniques}
Each type of malware has its own unique way of causing havoc, and most rely on user action of some kind. Some strains are delivered over email via a link or executable file. Others are delivered via instant messaging or social media. Even mobile phones are vulnerable to attack. It is essential that organizations are aware of all vulnerabilities so they can lay down an effective line of defense[22].
\subsubsection{Repackaging}
Repackaging includes the disassembling of the popular benign applications, then appending the malicious content and finally reassembling and distributing them on other less monitored third party markets. This is done by reverse-engineering tools.
\subsubsection{Drive By Download}
It Occurs when a user visits a website that contains malicious content and downloads malware into the device.
\subsubsection{Dynamic Payloads}
Uses dynamic payload to download an embedded encrypted source in an application. After installation, the application decrypts the encrypted malicious payload and executes the malicious code.
\subsubsection{Stealth Malware Technique}
Stealth Malware Technique refers to an exploit of hardware vulnerabilities to obfuscate the malicious code to easily bypass the anti-malware.

\subsection{Malware Evasion Techniques}
\subsubsection{Anti-security technique} 
These techniques are used to avoid detection by security devices and programs as anti-malwares, malwares, firewalls, and any other tools that protect the environment.
\level{4}{Fragmentation}
The malware splits into several components that only execute when it is reassembled.
\subsubsection{Anti-sandbox technique}
Anti-sandbox technique is used to detect automatic analysis and to avoid reports on the behavior of malware. This can be done by detecting registry keys, files or processes related to virtual environments environments.
\level{4}{Stalling Delays}
The malware simply does nothing for an extended period. Typically, 10 minutes is sufficient for most sandboxes[52] to timeout and assume the object is benign.
\level{4}{Suspended Activities}
The malware postpones these malicious actions while it is operating within a sandbox.
\newline A) Injection or modification of code within other applications.
\newline B) Establish persistence and download additional code.
\newline C)Move laterally across the network.
\newline D) Connect to its C\&C servers.
\level{4}{Rootkits} The malware hides malicious code in the lower layers of the operating system where conventional sandbox technology can’t see it.
\subsubsection{Anti-analyst techniques}
In these techniques, a monitoring tool is used to avoid reverse engineering. The tools might be process explorer or Wireshark to perform monitoring and to detect malware analysts[43]. 
\level{4}{ROP Evasion} Return-Oriented Programming (ROP) The malware injects functionality into another process without altering the code of that process. This is achieved by modifying the contents of the stack, which is the set of memory addresses that tell the system which segment of code to execute next.
\level{4}{User Action Required} The malware avoids doing anything malicious until a user performs a specific action (e.g., a mouse click, pressing a key, opening or closing a file, or exiting the program).
\newline

Malware creators might use two or three of the above techniques to make detection more difficult[43][44].
\newline
The rest of the paper is structured as  follows. Section(2) gives detailed view of the different type of malware detection Techniques, section(3) contain the details of tools for malware detection, section(4) is about the datasets for malware detection, section(5) contain the comparison of different malware detection studies, section(6) has some challenges from previous studies, section(7) contains the conclusion and the last section(8) contains the References used in this paper.All the tables used are at the end of paper.

\section{ Malware Detection}
Malware detection has multiple stages, which work together to detect or classify the malware. All the previous study focuses on malware detection in windows, smartphones and embedded systems(IOT) mainly. The study on malware detection is increasing in smartphones nowadays. The method for detection of malware is changing day by day as new researches come based on the increasing complexity of malware. The main malware detection process remains mostly same for all the studies as following[1]:
\newline 1) Malware analysis
\newline 2) Feature Extraction/selection
\newline 3)classification/detection
\newline
There are mainly 2 types of malware analysis static method and dynamic method which are mainly used to analyse the malicious file based on various parameters. Static Malware Analysis (SMA)[34]  where only basic analysis is done and malwares are detected without executing them. The methods used for static analysis are Basic Information Analysis, Structure Analysis, and Control Flow Analysis etc. But malwares which uses different measures of Polymorphism, Metamorphism, ShellCode etc. Can not be detected by static analysis.Some of the tools[12] for static analysis are in table 1. Dynamic Malware Analysis (DMA)[34] is done at the time of program execution. This technique is useful in analyzing the malwares which uses techniques such as Polymorphism, Metamorphism, Shell Code etc. But it is not useful in detecting Zero day malwares. Some of the tools for dynamic analysis are in table 1. Hybrid Analysis [34] is proposed to overcome the limitations of static and dynamic analysis techniques. It firstly analyses the signature specification of any malware code and then combines it with the other behavioral parameters for enhancement of complete malware analysis. Due to this approach hybrid analysis overcomes the limitations of both static and dynamic analysis. For detecting the malicious file for each technique we need some features which will become the inputs to the detection process. But a file contains a large number of features and  not all features are beneficial for a particular detection/classification algorithm and a large number of features can increase the execution time so we need to extract the correct features based on requirement. Sometimes by just changing the feature selection criteria we can increase accuracy. Sefer Kurnaz and Mokhalad Eesee Khudhur[18] used 4 data mining classification algorithms to classify the malware files. They used SVM, Random Forest, KNN and Hoeffding Tree over the dataset[for win32] containing 12593 malicious and 2405 benign files and they have used weka tool and three sets of features to pass to data mining algorithms. According to them random forest gives the best results[accuracy 98\%] to classify the malicious files based on different parameters like TPR, FPR, Accuracy, Recall, Precision and Receiver Operating characteristically graph. They claim that with the same algorithm and dataset their system give better accuracy than previous study by changing the feature selection methods.They used three different sets of feature selection methods "Symmetrical Uncertainty AttributeEval", "Information Gain (IG)", and " Correlation Attribute Eval ", as the best criterion utilized to select best features and the accuracy by the same algorithm increased.

\begin{figure*}[htpb!] % Defines figure environment
    \centering % Centers your figure
\includegraphics[width =18cm,height=20cm]{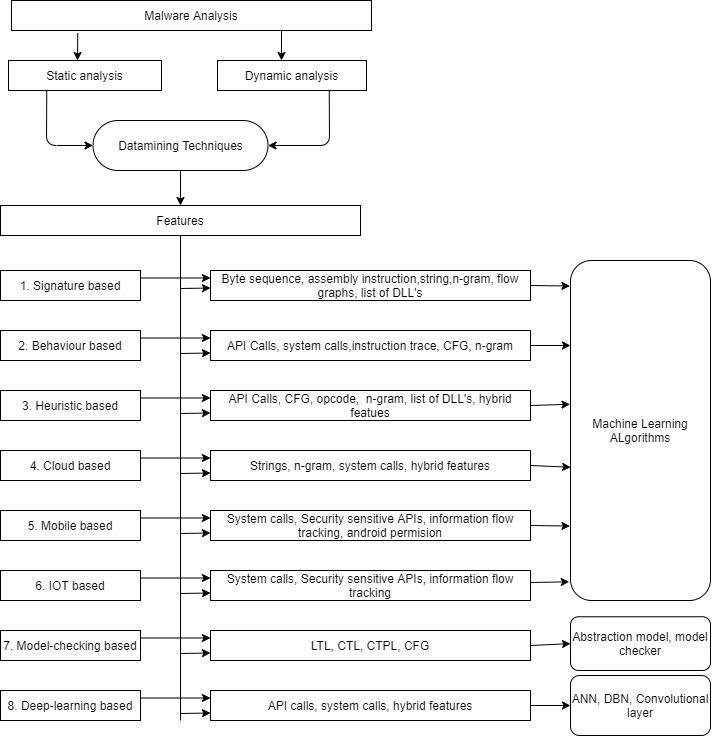} % Includes your figure and defines the size
    \caption{Malware detection Techniques} % For your caption
    \label{fig:label2} % If you want to label your figure for in-text references
\end{figure*}

\subsection{Malware Detection techniques}
In recent years the studies on malware detection has increased. The most used detection techniques in past are the signature based detection and the behaviour based detection and some studies used them both in combination as some feature taken from signatures detection and some are from behaviour of application which is under heuristic or hybrid detection. The subsection(2.1.3) discuss heuristic approach in detail. Nowday studies of malware detection focus mostly on mobile devices as the smartphones are the most used devices and most vulnerable. At the back all of the studies used machine learning approaches[1]. But we can classify them according to the platform and detection methodology in different form as deep learning based detection (which is part of machine learning but it mainly focus on neural networks, so we will discuss them in details), cloud based (as the detection are done somewhere at remote servers), IOT based(malware in embedded systems) and many more as shown in figure~\ref{fig:label2}.

\subsubsection{Signature-based detection}
Nowadays pattern matching is the most common method in malware detection, and signature based detection is the most popular method in this area [35]. Signature is a unique feature for each file, something like a fingerprint of an executable. Signature based methods use the patterns extracted from various malwares to identify them and are more efficient and faster than any other methods. These signatures are often extracted with special sensitivity for being unique, so those detection methods that use this signature have small error rates. Where this small error rate is the main reason that most common commercial antiviruses use this technique. These methods are unable to detect unknown malware variants and also require a high amount of manpower, time, and money to extract unique signatures. These are the main disadvantages of these methods. Also, inability to confront against the malwares that mutate their codes in each infection such as polymorphic and metamorphic one is another disadvantage. To tackle these challenges, research societies propose completely new malware detection families. It can not detect unknown and polymorphic malware variants.\newline
Ömer ASLAN [29] has compared the Static Malware Analysis Tools and Antivirus Scanners To Detect Malware and  shows that it is difficult to detect malware by only using one static tool or a few tools. Using only static analysis tools or antivirus software may not be enough as well. To correctly mark a suspicious program, it is recommended to use static tools with antivirus scanners. For unknown malware, the performance of the antivirus software declined sharply. The detection rate declined from 79\% to 56\% and accuracy declined from 80\% to 65\%. These results show that antivirus software cannot detect zero-day malware. Signature-based detection tools such as antivirus scanners are fast and effective when detecting existing malware, but it is almost impossible to detect unknown malware. On the other hand, static detection tools are more accurate when detecting more complex and zero-day malware. However, static analysis tools cannot detect a lot of new unknown malware too.

\subsubsection{Behavior-based detection}
Behavior based malware detection techniques observe behavior of a program to conclude whether it is malicious or not [35]. Since behavior based techniques observe what an executable file does, they are not susceptible to the shortcomings of signature-based ones. Simply put, a behavior based detector concludes whether a program is malicious by inspecting what it does rather than what it says. In these methods, programs with the same behavior are collected. Thus, a single behavior signature can identify various samples of malware. These types of detection mechanisms help in detecting malware that keep on generating new mutants since they will always use the system resources and services in the similar manner. A behavior-based detector basically consists of the following components[35]:
\begin{enumerate}[noitemsep]
    \item Data Collector: This component collects dynamic/static information about the executable. 
    \item Interpreter: This component converts raw information collected by data collection modules into intermediate representations. 
    \item Matcher: It is used to compare this representation with the behavior signatures. 
\end{enumerate}

There’s a multitude of behaviors that point to potential danger. Here are some examples:
\begin{itemize}[noitemsep]
    \item Any attempt to discover a sandbox environment
    \item Disabling anti-virus or other security controls
    \item Modifying the boot record or other initialization files to alter boot-up
    \item Installing rootkits
    \item Registering for autostart
    \item Shutting down or disabling system services
    \item Downloading and installing unknown software
    \item Deleting, altering, or adding system files
    \item Modifying other executable programs
    \item Connecting with known malicious sites
    \item Encrypting files that are unrelated to the program
    \item Adding or modifying user accounts
    \item Dynamic code building to enhance evasion capabilities
    \item Executing a dropped file
    \item Spawning Powershells
    \item Performing any actions that are highly abnormal
\end{itemize}
Disadvantage of Behaviour based detection:
\begin{itemize}[noitemsep]
    \item Non availability of promising false positive ration(FPR)
    \item High scanning time.
    \item Can not detect zero day malwares properly.
\end{itemize}

\subsubsection{Heuristic Based Detection:}
In heuristic method detectors use the features from both signature and behaviour technique and use that combined to detect the malware which changes depending upon various things. It uses features like API calls, CFG, opcode, n-gram, list of DLLs and other hybrid features. At the back this technique can use any machine learning algorithm to train and test the model and classify or detect malware. Although it has a high accuracy rate to detect zero-day malware to a certain degree, it cannot detect complicated malware. To overcome the disadvantages of signature, and behavioral-based malware detection approaches Z. Bazrafshan[35] gives the survey on heuristic detection methods and machine learning algorithms used. He gives detailed study of  the features API Calls, CFG, N-gram, Opcode and hybrid features. He used a machine learning algorithm to generate a pattern which was similar to signature. Based on the signature, new suspicious programs were marked malware or benign.

\subsubsection{Deep learning Based Detection}
Deep learning is part/subfield of artificial neural networks(ANN) and able to learn without human supervision, drawing from data that is both unstructured and unlabeled. This is mainly used to reduce features in malware detection.
Berman[14]gives the detailed view of various neural networks used in malware detection like deep belief network, Recurrent  NN, Convolutional NN,Generative adversarial network, Recursive NN and various open datasets with 2k to 4M files. Alzaylaee[2] proposed DL-Droid an application using dynamic analysis and stateful input generation in detection of malware in android and compared the detection performance and code coverage of the stateful input generation method with the commonly used stateless approach using the deep learning system. The DL-Droid gives 97.8\% detection rate (with dynamic features only) and 99.6\% detection rate (with both dynamic and static features) respectively which outperforms traditional machine learning techniques. It runs the applications on real mobile phones so that more accuracy can be achieved. It used the Dynalog dynamic analysis framework[31]. Zhongru Ren[10] proposed End-to-end malware detection for android IoT devices using deep learning. This methods resample the raw bytecodes of the classes.dex files of Android applications as input to deep learning models.Dataset contains 8K benign applications and 8K malicious applications from play store[46] and virusshare[47]and Prepare two models, first model called DexCNN with 93.4\% detection accuracy and the second model called DexCRNN can achieve a detection accuracy of 95.8\%. Heba Ziad Alawneh[15] proposes a dynamic malware detection approach for android applications. They use data mining over process execution time and extract process control block(PCB) information and apply a combination of CNNs, LSTM, and DNNs on it to identify the malicious application.

\subsubsection{IOT  and Mobile based detection}
Internet of Things (IoT) devices refers to Internet-connected smart devices such as home appliances, network cameras, and sensors. The IoT and mobile devices are being used more than PCs. Since mobile and IoT devices are becoming more popular among users day by day, they are also becoming more favorite targets for attackers. Because of that the malware detection schema landscape is changing from computers to IoT and mobile devices. Andrei Costin[19] gives the analysis of all currently known IoT malware families and uploaded this work as open-source material[59][60]. Xu Jiang[5] proposed Android Malware Detection Using Fine-Grained Features, which uses the permission used by application as static feature and evaluates 1700 benign applications and 1600 malicious applications and achieves a TP rate of 94.5\%. Author claims FDP can detect more malware families and only requires 15.205 s to analyze one application on average. Moutaz[4] proposed a system for classifying mobile applications using real-world datasets and applied two feature selection methods, Chi - Square and ANOVA with 10 supervised ML algorithms and achieved 98.1\%detection accuracy with a classification time of 1.22s on an average application. Limin Shen[3] proposes an application behavior- detection method based on multi feature and process algebra for detecting privilege escalation attacks in Android applications. By analysis of the privilege escalation attack model, five features are extracted. Attack model and application behaviour is built using process algebra. Dataflow path detection is conducted among the applications to determine those apps constituted a privilege escalation attack and  DroidBench benchmark test is used to test the accuracy and effectiveness of the proposed method. 
\subsubsection{Model Checking based detection}
In this detection approach, malware behaviors are manually extracted and behavior groups are coded using linear temporal logic (LTL) to display a specific feature[1]. Program behaviors are created by looking at the flow relationship of one or more system calls and define behaviors by using properties such as hiding, spreading, and injecting. By comparing these behaviors, it is determined whether the program is malware or benign. Model checking-based detection can detect some new malware to a certain degree, but cannot detect all new generations of malware. This is a very old Technique.Kinder et al. proposed a flexible method to detect malicious code patterns in executables by model checking [40]. They introduced the specification language CTPL (computation tree predicate logic) which extends the well-known logic CTL (computation tree logic), and describes an efficient model checking algorithm. According to the authors, test results demonstrated that the proposed method can detect many worm variants with a single specification. Proposed method has some limitations as, It Can only detect worm variants, Some part of the process has to be done manually, Performance of proposed method is low. To get better results, CTPL can be extended to detect other malware. In addition, more accurate data integrity constructions and efficient data structures can be used to improve the method performance.

\subsubsection{Cloud based detection}
Cloud computing is a very growing technology and now can be used in the malware detection field by using security as a service. Users can upload any file and get the result as it is malicious or not. Cloud has capacity to store large dataset so it can enhance the detection performance of any pc or mobile by security as a service. Martignoni presented a framework that enhances the capabilities of existing dynamic behavior-based detectors. The proposed framework enables sophisticated behavior based analysis of suspicious programs in multiple realistic and heterogeneous environments in the cloud [38]. The suggested schema forces sample programs to execute in a distributed environment including security labs and potential victim machines. The evaluation results demonstrated that the analysis of multiple execution traces of the same malware  sample in multiple end-users’ environments can improve the results of the analysis with very small overhead. On the other hand, the suggested framework raises the privacy and security issues, and is prone to various forms of detection and evasion attacks. Solving security related issues and implementing a resistant framework against evasion attacks will increase the framework performance. Andrew McDole, Mahmoud[13] analyzes and compares various Convolutional Neural Networks (CNNs) for online detection of malware in cloud IaaS. They analyzed seven different convolutional neural network models and determined which model is better suited for malware detection in cloud IaaS. The analysis shows that the LeNet-5 model is quick but gives less accuracy. This model gives 90\% accuracy and can be used in situations where  incorrectness is not too costly but a quick prediction is needed. Yanfang[36] presented a cloud-based schema to improve malware detection results by combining the file content and file relations and developed a file verdict system. The system incorporated into the Comodo’s Anti-malware products, and empirical studies were conducted on large daily datasets collected by Comodo cloud security center. The authors claims, the accuracy and efficiency of the Valkyrie system outperform other popular anti-malware software tools such as Kaspersky AntiVirus and McAfee VirusScan, as well as other alternative data mining based detection systems.

\subsubsection{Machine Learning Algorithms}
\begin{figure}[t]
    \centering
\includegraphics[scale=0.5]{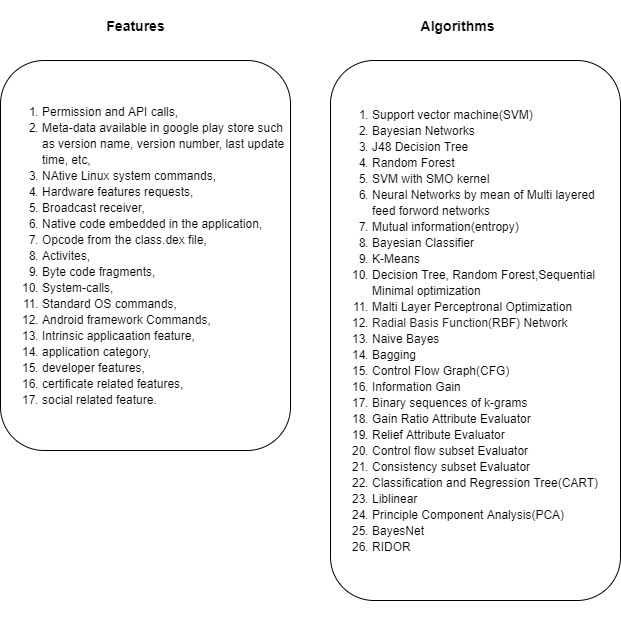} 
    \caption{Features and Machine Learning Algorithms for Static Analysis.}
    \label{fig:label3} 
\end{figure}

\begin{figure}[t]
    \centering 
\includegraphics[scale=0.5]{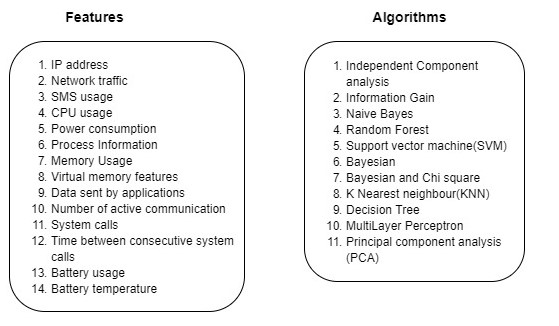}
    \caption{Features and Machine Learning Algorithms for Dynamic Analysis.} 
    \label{fig:label4} 
\end{figure}

Most detection uses different types of machine learning algorithms for classification and detection. Figures~\ref{fig:label3}~,\ref{fig:label4} shows the features and machine learning algorithms for static and dynamic analysis. For the malware detection we can use any features from figures~\ref{fig:label3}~,\ref{fig:label4} based on the type of file, can apply any machine learning algorithm and choose the best algorithm based upon accuracy, logloss or other performance measure as discussed in upcoming section 5. For example if we have an android(.apk) file and we want to classify whether it is malicious or legitimate, we can extract permission, api call, or system call feature and can apply any classification algorithm like SVM, Random forest etc.

The following are the factors which can affect the machine learning algorithm:
\begin{enumerate}[noitemsep]
    \item Dataset
    \item Type of features
    \item Feature-selection algorithm used to select the most prominent features
    \item Classification algorithm used to categorize apps as malicious or clean
    \item Classifier’s parameter values(Hyperparameters)
\end{enumerate}

\section{Tools for Malware Detection}
Table~\ref{tab:tools} shows some tools that are available for detection and analysis.

\section{ Datasets for Malware Detection}
Some frequently used standard datasets and datset repositories by researchers are listed in table~\ref{tab:dataset}.

\section{Comparison of Malware Detection Techniques}
The researchers used different algorithms for malware detection. Most of the studies uses machine learning techniques to identify the malware. Table~\ref{tab:comparision} shows the comparison of different studies in malware detection with the malware detection methods used and the accuracy of each studies. Tables~\ref{tab:compare} gives a good comparison of malware detection techniques, we have discussed in section 2. 

\subsection{Performance Measures}
Some of the matrices for identifying how good the classification (performance measures) is are following:
\begin{itemize}[noitemsep]
    \item Accuracy \item Precision \item Recall \item False positive ratio \item F1 score \item Area under curve(AUC) \item Log loss
\end{itemize}

These matrices can be derived from:
\begin{itemize}[noitemsep]
    \item True positive(TP)  \item False positive(FP)   \item True negative(TN)  \item False negative(FN)
 \end{itemize}

\section{Challenges }
The Followings are some challenges in malware detection.
\begin{enumerate}[noitemsep]
    \item A little information for classification.
    \item Comparing a few algorithms.
    \item High scanning/detecting time.
    \item High FP and FN.
    \item Small Datasets.
    \item Overfitting.
    \item Can not detect zero day malwares properly.
    \item Not detecting Hidden Malwares.
    \item Analysis is done on Datasets instead of Real Time monitoring.
\end{enumerate}

\section{Conclusion}
We have summarized 8 detection methods for malware, but no method is completely able to detect all new generation malware. Only the behaviour based and the model checking based detection can resist the obfuscation. According to the discussion the deep learning based and the cloud based also can detect the malware very well but they are not able to detect all types of malware completely. As a future work new method and work are needed. This paper will help to understand the techniques available for malware detection till date and it can be used as a good reference for the further studies.

\section{References}
\begin{enumerate}[noitemsep]
    \item Ö. A. Aslan and R. Samet, "A Comprehensive Review on Malware Detection Approaches," in IEEE Access, vol. 8, pp. 6249-6271, 2020, doi: 10.1109/ACCESS.2019.2963724.
\item Alzaylaee, Mohammed K., Suleiman Y. Yerima, and Sakir Sezer. "DL-Droid: Deep learning based android malware detection using real devices." Computers \& Security 89 (2020): 101663.
\item Limin Shen, Hui Li, Hongyi Wang, Yihuan Wang, "Multifeature-Based Behavior of Privilege Escalation Attack Detection Method for Android Applications", Mobile Information Systems, vol. 2020, Article ID 3407437, 16 pages, 2020. https://doi.org/10.1155/2020/3407437
\item Alazab, Moutaz. "Automated Malware Detection in Mobile App Stores Based on Robust Feature Generation." Electronics 9.3 (2020): 435.
\item Xu Jiang, Baolei Mao, Jun Guan, Xingli Huang, "Android Malware Detection Using Fine-Grained Features", Scientific Programming, vol. 2020, Article ID 5190138, 13 pages, 2020. https://doi.org/10.1155/2020/5190138
\item Alazab, Moutaz, et al. "Intelligent mobile malware detection using permission requests and api calls." Future Generation Computer Systems 107 (2020): 509-521.
\item Y. A. Ahmed, B. Koçer and B. A. S. Al-rimy, "Automated Analysis Approach for the Detection of High Survivable Ransomware," KSII Transactions on Internet and Information Systems, vol. 14, no. 5, pp. 2236-2257, 2020. DOI: 10.3837/tiis.2020.05.021
\item Ali, Abdullah; Eshete, Birhanu; (2020). Best-Effort Adversarial Approximation of Black-Box Malware Classifiers. arXiv preprint arXiv:2006.15725
\item Alper Egitmen, Irfan Bulut, R. Can Aygun, A. Bilge Gunduz, Omer Seyrekbasan, A. Gokhan Yavuz, "Combat Mobile Evasive Malware via Skip-Gram-Based Malware Detection", Security and Communication Networks, vol. 2020, Article ID 6726147, 10 pages, 2020. https://doi.org/10.1155/2020/6726147
\item Ren, Zhongru \& Wu, Haomin \& Ning, Qian \& Hussain, Iftikhar \& Chen, Bingcai. (2020). End-to-end Malware Detection for Android IoT Devices Using Deep Learning. Ad Hoc Networks. 101. 102098. 10.1016/j.adhoc.2020.102098. 
\item Sudhakar, Kumar, S. An emerging threat Fileless malware: a survey and research challenges. Cybersecur 3, 1 (2020). https://doi.org/10.1186/s42400-019-0043-x
\item Talukder, Sajedul. (2020). Tools and Techniques for Malware Detection and Analysis.https://arxiv.org/abs/2002.06819
\item McDole, Andrew, et al. "Analyzing CNN Based Behavioural Malware Detection Techniques on Cloud IaaS." arXiv preprint arXiv:2002.06383 (2020).
\item Berman, D.S.; Buczak, A.L.; Chavis, J.S.; Corbett, C.L. A Survey of Deep Learning Methods for Cyber Security. Information 2019, 10, 122.
\item Fei Xiao, Zhaowen Lin, Yi Sun, Yan Ma, "Malware Detection Based on Deep Learning of Behavior Graphs", Mathematical Problems in Engineering, vol. 2019, Article ID 8195395, 10 pages, 2019. https://doi.org/10.1155/2019/8195395
\item lawneh, Heba \& Umphress, David \& Skjellum, Anthony. (2019). Android Malware Detection Using Neural Networks \& Process Control Block Information. 
\item Rahim Taheri, Meysam Ghahramani, Reza Javidan, Mohammad Shojafar, Zahra Pooranian, Mauro Conti,Similarity-based Android malware detection using Hamming distance of static binary features, Future Generation Computer Systems, Volume 105, 2020, https://doi.org/10.1016/j.future.2019.11.034
\item Kurnaz, S. and Mokhalad Eesee Khudhur. “Comparative and Analysis Study for Malicious Executable by Using Various Classification Algorithms.” (2018).
\item Costin, Andrei. “IoT Malware : Comprehensive Survey , Analysis Framework and Case Studies.” (2018).
\item S. Anderson and P. Roth, ‘‘EMBER: An open dataset for training static PE malware machine learning models,’’ 2018 https://arxiv.org/abs/1804.04637
\item Pandey, Anjana. (2018). A STUDY ON DIGITAL FORENSICS USING  VARIOUS ALGORITHMS FOR MALWARE DETECTION. International  Journal of Advanced Research in Computer Science. 9. 85-89.  10.26483/ijarcs.v9i3.6084.  
\item Amro, Belal. (2017). Malware Detection Techniques for Mobile Devices. International Journal of Mobile Network Communications \& Telematics. 7. 10.5121/ijmnct.2017.7601. 
\item Nikola Milosevic a , Ali Dehghantanha b and Kim-Kwang Raymond Choo, "Machine learning aided Android malware classification," Computers \& Electrical Engineering, vol. 61, pp. 266-274, 2017
\item A. H. Lashkari, A. F. A. Kadir, H. Gonzalez, K. F. Mbah, and A. A. Ghorbani, ‘‘Towards a network–based framework for Android malware detection and characterization,’’ in Proc. 15th Annu. Conf.Privacy,Secur. Trust (PST), Aug. 2017
\item Bat-Erdene M, Park H, Li H, Lee H, Choi MS (2017) Entropy analysis to classify unknown packing algorithms for malware detection.Int J Inf Secure 16(3):227–248. https://doi.org/10.1007/s10207-016-0330-4
\item Narayanan A, Chandramohan M, Chen L, Liu Y (2017) A multi-view context-aware approach to Android malware detection and malicious code localization. Empir Softw Eng. https://doi.org/10.1007/s10664-017-9539-8
\item Alam S, Qu Z, Riley R, Chen Y, Rastogi V (2017) DroidNative: automating and optimizing detection of Android native code malware variants. Comput Secur 65:230–246. https://doi.org/10.1016/j.cose.2016.11.011
\item Bhattacharya, Abhishek \& Goswami, Radha. (2016). DMDAM: Data Mining Based Detection of Android Malware.10.1007/978-981-10-2035-3$\backslash$\_20. 
\item Aslan, Ömer. (2017). Performance Comparison of Static Malware Analysis Tools  Versus Antivirus Scanners To Detect Malware.  
\item Ethan M. Rudd, Andras Rozsa, Manuel Günther, and Terrance E. Boult.A Survey of Stealth Malware: Attacks, Mitigation Measures, and Steps Toward Autonomous Open World Solutions.https://arxiv.org/abs/1603.06028v2.
\item Alzaylaee, M.K., Yerima, S.Y., and Sezer, S., 2016. Dynalog: an automated  dynamic analysis framework for characterizing android applications. In: 2016  International Conference On Cyber Security And Protection Of Digital Services  (Cyber Security), pp. 1–8. doi: 10.1109/CyberSecPODS.2016.7502337 . 
\item S. Alam, R. Horspool, I. Traore, and I. Sogukpinar, ‘‘A framework for metamorphic malware analysis and real-time detection,’’ Comput. Secur., vol. 48, pp. 212–233, Feb. 2015
\item S. K. Pandey and B. M. Mehtre, "Performance of malware detection tools: A comparison," 2014 IEEE International Conference on Advanced Communications, Control and Computing Technologies, Ramanathapuram, 2014, pp. 1811-1817, doi: 10.1109/ICACCCT.2014.7019422.
\item Jyoti Landage, Prof. M. P. Wankhade, 2013, Malware and Malware Detection Techniques : A Survey, INTERNATIONAL JOURNAL OF ENGINEERING RESEARCH \& TECHNOLOGY (IJERT) Volume 02, Issue 12 (December 2013),
\item Z. Bazrafshan, H. Hashemi, S. M. H. Fard and A. Hamzeh, "A survey on heuristic malware detection techniques," The 5th Conference on Information and Knowledge Technology, Shiraz, 2013, pp. 113-120, doi: 10.1109/IKT.2013.6620049
\item Y. Ye, T. Li, S. Zhu, W. Zhuang, E. Tas, U. Gupta, and M. Abdulhayoglu, ‘‘Combining file content and file relations for cloud based malware detection,’’ in Proc. 17th ACM SIGKDD Int. Conf. Knowl. Discovery Data Mining (KDD), 2011
\item kdd(M. Tavallaee, ‘‘A detailed analysis of the KDD CUP 99 data set,’’ in Proc.IEEE Symp. Comput. Intell. Secur. Defense Appl., 2009, pp. 1–6.)
\item L. Martignoni, R. Paleari, and D. Bruschi, ‘‘A framework for behavior based malware analysis in the cloud,’’ in Proc. Int. Conf. Inf. Syst. Secur. Berlin, Germany: Springer, 2009
\item Z. Zuo, Q. Zhu, and M. Zhou, ‘‘On the time complexity of computer viruses,’’ IEEE Trans. Inf. Theory, vol. 51, no. 8, pp. 2962–2966,Aug. 2005
\item J. Kinder, S. Katzenbeisser, C. Schallhart, and H. Veith, ‘‘Detecting malicious code by model checking,’’ in Proc. Int. Conf. Detection Intrusions Malware, Vulnerability Assessment. Berlin, Germany: Springer, 2005
\item Gergely Erdelyi, "Hide’n’Seek Anatomy of stealth malware” https://www.blackhat.com/presentations/bh-europe-04/bh-eu-04-erdelyi/bh-eu-04-erdelyi.pdf
\item D. Spinellis, ‘‘Reliable identification of bounded-length viruses is NP-complete,’’ IEEE Trans. Inf. Theory, vol. 49, no. 1, pp. 280–284,Jan. 2003
\item https://www.csoonline.com/article/2615925/security-your-quick-guide-to-malware-types.html
\item https://www.lastline.com/understanding-advanced-threat-malware-detection
\item https://www.techopedia.com/definition/4130/stealth-virus
\item https://play.google.com/
\item https://virusshare.com/
\item https://www.virustotal.com.
\item http://www.malgenomeproject.org/
\item https://www.comodo.com/home/internet-security/updates/vdp/database.php
\item http://contagiodump.blogspot.com/
\item https://app.sndbox.com/
\item https://www.sec.cs.tu-bs.de/~danarp/drebin/
\item http://nlp.cs.aueb.gr/software\_and\_datasets/Enron-Spam/index.html
\item https://spamassassin.apache.org/
\item http://arxiv.org/abs/1802.10135
\item https://kilthub.cmu.edu/articles/dataset/\newline Insider\_Threat\_Test\_Dataset/12841247/1
\item https://androzoo.uni.lu/
\item http://firmware.re/malw
\item http://firmware.re/bh18us
\end{enumerate}

\begin{table*}
    \centering
    \scriptsize
\caption{Tools for malware detection}
   \begin{tabular}{|m{0.2\linewidth} | m{0.3\linewidth}| m{0.5\linewidth}| } 
\hline
Tool Type & Tool Name & Description \\
\hline
\multirow{5}{4em}{Detection Tools} & Analyse PE & Wrapper for a variety of tools for reporting on window PE files.\\\cline{2-3} 
& CHKrootkit  & Linux rootkit detector.\\\cline{2-3} 
& MASTIFF & Static analysis Framework. \\\cline{2-3}
& MultiScanner & Modular file scanner.\\\cline{2-3}
& PEV & for analysis of suspicious binar\\
\hline

\multirow{4}{4em}{Online Scanner and Sandbox} & Andro Total &
Online analysis of APKs against multiple mobile antivirus apps.\\\cline{2-3}
& APK Analyzer & Dynamic analysis of APKs.\\\cline{2-3}
& Cuckoo sandbox & Open source sandbox and automated analysis system.\\\cline{2-3}
& Deepviz & Multiformat file analyzer with machine learning classifier. \\

\hline
\multirow{6}{4em}{Static Analysis Tools} & PEid  & Detects most common packers, crypters and compilers for PE files.\\\cline{2-3}
 & Resource Hacker & Used to add, modify or replace most resources within Windows binaries including strings, images, dialogs, menus, VersionInfo and Manifest resources.\\\cline{2-3}
  & Dependency walker & List the imported and exported functions of a PE file. It also displays a recursive tree of all the dependencies of the executable file.\\\cline{2-3}
 & PEView. & Provides a quick and easy way to view the structure and content of 32-bit PE and COFF files.\\\cline{2-3}
 & apktool & Decompile the applications.\\\cline{2-3}
 & IDA PRO & Disassembler to generates assembly language source code from machine-executable code. \\
 
 \hline

\multirow{5}{4em}{Dynamic  Analysis Tools}  & Rogshot & Capture a snapshot of the system prior to executing malware and then immediately afterwards.\\\cline{2-3}
 & Process Explorer. & Give  real-time system information about the running process.\\\cline{2-3}
 & Process monitor & Realtime troubleshooting tool.\\\cline{2-3}
  & Immunity debugger  & Write exploits, analyze malware, and reverse engineer binary files.\\\cline{2-3}
 & ollyDbg & Traces registers, recognizes procedures, API calls, switches, tables, constants and strings, as well as locates routines from object files and libraries. \\

\hline

\end{tabular}
\label{tab:tools}
\end{table*}

\begin{table*}
    \centering
    \scriptsize
    
    \caption{Datasets for malware detection}
    \begin{tabular}[t]{|m{0.04\linewidth} | m{0.4\linewidth}| m{0.4\linewidth}|}
        \hline Sr.No. & Dataset & Description \\
         \hline 1. & Knowledge discovery and dissemination(KDD) 199 dataset[37]  &  Approximately 4,900,000 single connection vectors, each  contains 41 features. \\
         \hline 2.&Genome Project dataset[49]
 & 1,200 malware samples.\\
        \hline 3.& Virusshare[46]& 106,555 bytes different datasets for various malware families. \\
        \hline 4.&VirusTotal dataset[48] & 
Provide 70 antivirus scanner and URL/Domain blacklisting services. Need to upload files to check if it is malicious or not. \\
        \hline 5.& Comodo dataset[50]& 79666064 file till 2/12/20, Updates in every 2 days. \\
        \hline 6.&Contigeio dataset[51] &  189 malware samples.\\
        \hline 7.& DREBIN dataset[54]& 5,560 applications from 179 different malware families. \\
        \hline 8.& Microsoft dataset[56] & Dataset is almost half a terabyte, malware files representing a mix of 9 different families. \\
        \hline 9.& CERT  insider threat dataset v6.2[5][57]&  Contain multiple dataset over 83 gb files.\\
        \hline 10 & EnronSpam[55] & 30207 emails of which 16545 emails are labeled as ham and 13662 emails are labeled as spam.\\
        \hline 11 & SpamAssassin[55] & 6047 messages, with  31\% spam ratio. \\
        \hline 12.& LingSpam[14] &Open Source(for Email Spam Check) 2,893 spam and non-spam messages.  \\
        \hline 13.& SNDBOX[53] & Free open source (200 MB of files). \\
        \hline 14.&Ember[20] &  500MB, consisting of disassembly and byte- code of around 20K malicious samples from nine families.(Open source).\\
        \hline 15.&Androzoo[58] &  Contains 13,996,153 different APKs, each of which has been analysed by tens of different AntiVirus products to know which applications are detected as Malware.\\
        \hline
    \end{tabular}
    \label{tab:dataset}
\end{table*}

\begin{table*}
\centering
\scriptsize
      \caption{Comparison of malware detection Studies}
\begin{tabular}[b]{| m{0.08\linewidth} | m{0.3\linewidth}| m{0.3\linewidth} | m{0.3\linewidth} |} 

    \hline S.No. & Name & Method & Performance Measures \\
    \hline 1. & Static analysis tools vs antivirus scanner[29] & Signature based & For static -max 68.2\% \newline For antivirus scanner-max 58.9\% \\
    
    \hline 2.  & Digital Forensic- comparing different algorithm[21] & Machine Learning  &FPR- 0.417\% \newline
FNP - 0.716802\% \newline
GNB - 70.18\% \newline
 DecisionTree-99.11\% \newline R.Forest - 99.4929\% \newline AdaBoost- 98.4534\% \newline  G.Boosting-98.801\%  \\
    
 \hline 3.  & DL-Droid Deep learning framework[2] & Deep Learning  &  97.8\%  (with dynamic features only) \newline99.6\% (with dynamic + static features)  \\ 

\hline 4. & End-to-end Md for android IOT[10] proposed 2 models & Deep Learning & DexCNN -93.4\% \newline  DexCRNN - 95.8\%.  \\

\hline 5.  & Analyzing CNN Based Behavioural Malware Detection Techniques on Cloud IaaS[13] & Cloud based detection & LeNet5 - 89.9\% \newline  
ResNet50 - 90.7\% \newline 
ResNet101 - 87.0\% \newline  
ResNet152 - 88.7\% \newline DenseNet121 - 92.1\% \newline  
DenseNet169 - 91.9\% \newline 
DenseNet201 - 91.5 \\

\hline 6. & Comparative and Analysis Study for Malicious Executable by Using Various Classification Algorithms[18] & Machine Learning & SVM - 96.12\% \newline KNN- 97.87\% \newline 
Ho-effding Tree- 94.5\% \newline
Random forest-98.12\% \\

\hline 7. & A framework for metamorphic malware analysis and real-time detection[32] & ACFG (Annotated Control Flow Graph) and SWOD (Sliding Window of Difference and CFG)
 - CFWeight & In range 94\% - 99.6\% \\

\hline 8. & Android Malware Detection Using Fine-Grained Features[5] & Machine learning & TP  -  94.5\%, 15.205 s time on an average application. \\

\hline 9. & Automated Malware Detection in Mobile App Stores Based on Robust Feature Generation[4] & IOT and Machine Learning  & Detection Accuracy - 98.1\% \newline Classification time- 1.22s on an average application. \\
\hline 10. & Intelligent mobile malware detection using permission requests and API calls.[6] & Mobile based detection & F-measure - 94.31\% \\
\hline
11. & Automated Analysis Approach for the Detection of High Survivable Ransomware.[7] & Machine Learning and Deep learning based & ROC curve of 0.987 \newline F Rate - 0.007 \\
\hline
12. & Black-Box Malware Classifiers [8] & Deep Learning based and machine learning & LGBM -0.88 \newline 
Decision Tree - 0.85 \newline 
  Random Forest - 0.87 \newline KNN - 0.91 \\
\hline
13. & Combat Mobile Evasive Malware via Skip-Gram-Based Malware Detection[9] & Deep Learning based and machine learning &
RF - 95.64\% on entire dataset, 95\% on evasive only samples.
\newline 
Svm-81.94\%\newline
Decision Tree- 92.04\% \newline 
Random Subspace - 94.86\% \newline
KNN - 94.48\%  \newline
For test set containing only zero day without including them in training set RF- 37.36\% \\
\hline
14. & 
Malware Detection Based on Deep Learning of Behavior Graphs[16] & 
Deep Learning based & 
Precision-  0.986\newline
 Recall -  0.992 \newline
F1-Score -  0.989 \\
\hline
15. & 
Similarity-based Android malware detection using Hamming distance of static binary features[17] & 
Deep Learning based & 
For all algorithm and datasets accuracy is greater than 90\% and and in some cases (i.e., considering API features) are more than 99\%.\\

\hline
16. & 
DroidNative: Automating and optimizing detection of Android native code malware variants[27] & 
Mobile based & 
 detection rate (DR) - 93.57\%\newline
 false positive rate - 2.7\% \\
\hline
17. & 
Machine learning aided malware classification of Android applications [23] & 
Mobile based & 
F-score - 95.1\%  \\
\hline
18. & Network-Based Framework for Android Malware Detection [24] & 
Mobile based & 
accuracy - 91.41\%\newline
false positive - 0.085 \\
\hline
19. & Entropy analysis to classify unknown packing algorithms for malware detection[25] & 
Machine Learning based & 
Accuracy of - 95.35\%\newline
precision -  94.13\% \\
\hline
20. & A Multi-view Context-aware Approach to Android Malware Detection and Malicious Code Localization[26]
 & 
Mobile based and machine based  & 
average recall - 94\%  \\
\hline
21. & DMDAM: Data Mining Based Detection of Android Malware[28] & 
Mobile and machine learning based & 
 TPR rate -  96.70\% ,\newline
Accuracy is up to 77.13\%,\newline
Highest F1 score is 0.8583. \\
\hline
  
\end{tabular}
    \label{tab:comparision}

\end{table*}

\begin{table*}
\centering
\small
\caption{Comparision of Malware Detection Techniques}
\begin{tabular}{|m{0.3\textwidth} | m{0.1\textwidth}| m{0.1\textwidth}|m{0.1\textwidth}| m{0.1\textwidth}|}
    
    \hline Malware Detection Technique &Detect Unknown Malware & Resistant to Obfucation & Well-Known Approach & New Approach \\
    
    \hline Signature Based & $\times$ & $\times$ & $\surd$ & $\times$\\
    \hline Behavior Based & $\surd$ & $\surd$ & $\surd$ & $\times$\\
    \hline Heuristic Based  & $\surd$ & $\times$ & $\surd$  & $\times$ \\
    \hline Model Checking Based & $\surd$ & $\surd$ & $\surd$  &$\times$\\
    \hline Deep Learning Based & $\surd$ & $\times$& $\times$ & $\surd$ \\
    \hline Cloud Based  & $\surd$ & $\times$& $\times$&$\surd$ \\
    \hline Mobile Based Detection & $\surd$ & $\times$& $\times$&$\surd$ \\
    \hline IOT Based Detection & $\surd$ & $\times$& $\times$&$\surd$ \\
    \hline
    
    \end{tabular}
    
\label{tab:compare}
\end{table*}

\end{document}